\documentclass[12pt]{iopart}

\usepackage{graphicx}
\usepackage{epsfig}
\usepackage{subfigure}
\usepackage{bm}

\begin{document}

\title{\bf A Training effect on electrical properties in nanoscale BiFeO$_3$}

\author {Sudipta Goswami} \address {Nanostructured Materials Division, CSIR-Central Glass and Ceramic Research Institute, Kolkata 700032, India}
\author {Dipten Bhattacharya}
\ead{dipten@cgcri.res.in} \address {Nanostructured Materials Division, CSIR-Central Glass and Ceramic Research Institute, Kolkata 700032, India}
\author {Wuxia Li} \address {Laboratory of Microfabrication, Institute of Physics, Chinese Academy of Sciences, Beijing 100190, P.R. China}
\author {Ajuan Cui} \address {Laboratory of Microfabrication, Institute of Physics, Chinese Academy of Sciences, Beijing 100190, P.R. China}
\author {QianQing Jiang} \address {Laboratory of Microfabrication, Institute of Physics, Chinese Academy of Sciences, Beijing 100190, P.R. China} 
\author {Chang-zhi Gu} \address {Laboratory of Microfabrication, Institute of Physics, Chinese Academy of Sciences, Beijing 100190, P.R. China}

\date{\today}

\begin{abstract}
We report our observation of the training effect on dc electrical properties in a nanochain of BiFeO$_3$ as a result of large scale migration of defects under combined influence of electric field and Joule heating. We show that an optimum number of cycles of electric field within the range zero to $\sim$1.0 MV/cm across a temperature range 80-300 K helps in reaching the stable state via a glass-transition-like process in the defect structure. Further treatment does not give rise to any substantial modification. We conclude that such a training effect is ubiquitous in pristine nanowires or chains of oxides and needs to be addressed for applications in nanoelectronic devices.
\end{abstract}

\pacs{75.80.+q, 75.75.+a, 77.80.-e}
\maketitle

\section{Introduction}
The stability and reliability of the nanowires or chains is a crucial issue for applications in nanoelectronic or spintronic devices \cite{Schmidt}. The fluctuations arising out of electric field induced defect migration and/or metastability of phases \cite{Buhrman,Tu}, influence the electrical, magnetic, and optical properties significantly \cite{Saito}. Several authors have already reported influence of electromigration on electrical properties in both metallic and oxide nanosized systems and have resorted to measurements such as relaxation of the conductivity, noise spectra \cite{Ghosh}, to even direct imaging of defect migration by transmission electron, scanning tunneling or atomic force microscopy \cite{Strachan,Braun,Li1}. While electromigration of defects in nanoscale interconnects or thin film surfaces leads to destabilization and even failure, it can also train the sample by healing the stress developed otherwise \cite{Tomar}. In this paper, we too observe an example of healing in an interesting training effect on the electrical properties of nanochains of BiFeO$_3$. BiFeO$_3$ happens to be a rare example of room temperature multiferroic compound with strong cross-coupling between ferroelectric and magnetic order parameters \cite{Scott}. While the bulk sample exhibits strong multiferroic coupling at room temperature \cite{Lebeugle}, it remains to be seen how the coupling varies as the particle size decreases from bulk down to nanoscale. Direct electrical measurements on nanoscale structures - nanoparticles, nanowires, nanorods, or nanochains - of BiFeO$_3$ under a magnetic field would provide authentic data about the multiferroic coupling. However, we discovered that direct electrical measurement on a pristine nanochain of BiFeO$_3$ does not yield reproducible and thus authentic data - they vary as the measurement is repeated again and again. Interestingly, the results eventually converge to a reproducible value which does not change further beyond an optimum number of measurement cycles. This is quite akin to the training effect observed in magnetic properties \cite{Biternas} such as exchange bias, coercivity, saturation magnetization etc where because of relaxation, the important parameters vary with the number of measurement cycles and then converge to a stable value. Accordingly, we call a similar observation in the case of the electrical properties of a nanochain of BiFeO$_3$ the training effect on the electrical properties. We further noticed that this training effect results from large scale migration of defects under the combined influence of inhomogeneous Joule heating and electric field. We show that following a treatment of electric field cycling within zero to $\sim$1.0 MV/cm across a temperature range 80-300 K, the nanochains of BiFeO$_3$ reach a stable equilibrium with no further change in dc conductivity in subsequent cycles. The kinetics of the defects, determined from dielectric relaxation spectroscopy, appears to be slowing down drastically from the pristine to the stabilized state, indicating the occurrence of a glass-transition-like phenomenon. The $\textit{in situ}$ atomic force microscopy (AFM) images recorded under a bias field on a pristine nanochain of BiFeO$_3$ offer direct evidence of large scale migration of defects leading to a substantial change in the microstructure.  

\begin{figure}[!h]
  \begin{center}
    \includegraphics[scale=0.60]{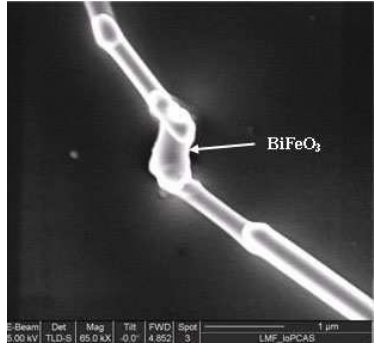} 
    \end{center}
  \caption{(a) A Scanning electron microscopic image of the BiFeO$_3$ nanochain patterned by focused ion beam in the two-probe configuration. }
\end{figure}

\section{Experiments}
The nanochains of BiFeO$_3$ system have been fabricated from solution chemistry route \cite{Goswami}. The chains have been patterned in a two-probe configuration with tungsten wires by focused ion beam (FIB) milling and deposition. In order to ensure a completely contamination-free connection, an ultra-low deposition current has been used (1-50 pA) and minor milling has been applied after the deposition of the wires. A representative scanning electron microscopic image of such a structure is shown in figure 1. The dimensions of the nanochain are: diameter $\sim$0.4 $\mu$m and length $\sim$0.8 $\mu$m. The measurement of dc current-voltage characteristics was carried out both in current and voltage driven mode. The HP high resistance meter 4339B and Keithley precision current source 6220 were used for the measurements. An Oxford Instrument cryostat were used for measuring the characteristics across 80-300 K. The dielectric relaxation properties were measured by Hioki LCR meter 3632-50 across a frequency range of 100 Hz - 2 MHz with a 1 V field amplitude. The $\textit{in situ}$ AFM images have been captured in contact mode  under a bias field ($\sim$0.125 MV/cm) at room temperature.

\begin{figure*}[htp]
  \begin{center}
    \subfigure[]{\includegraphics[scale=0.22]{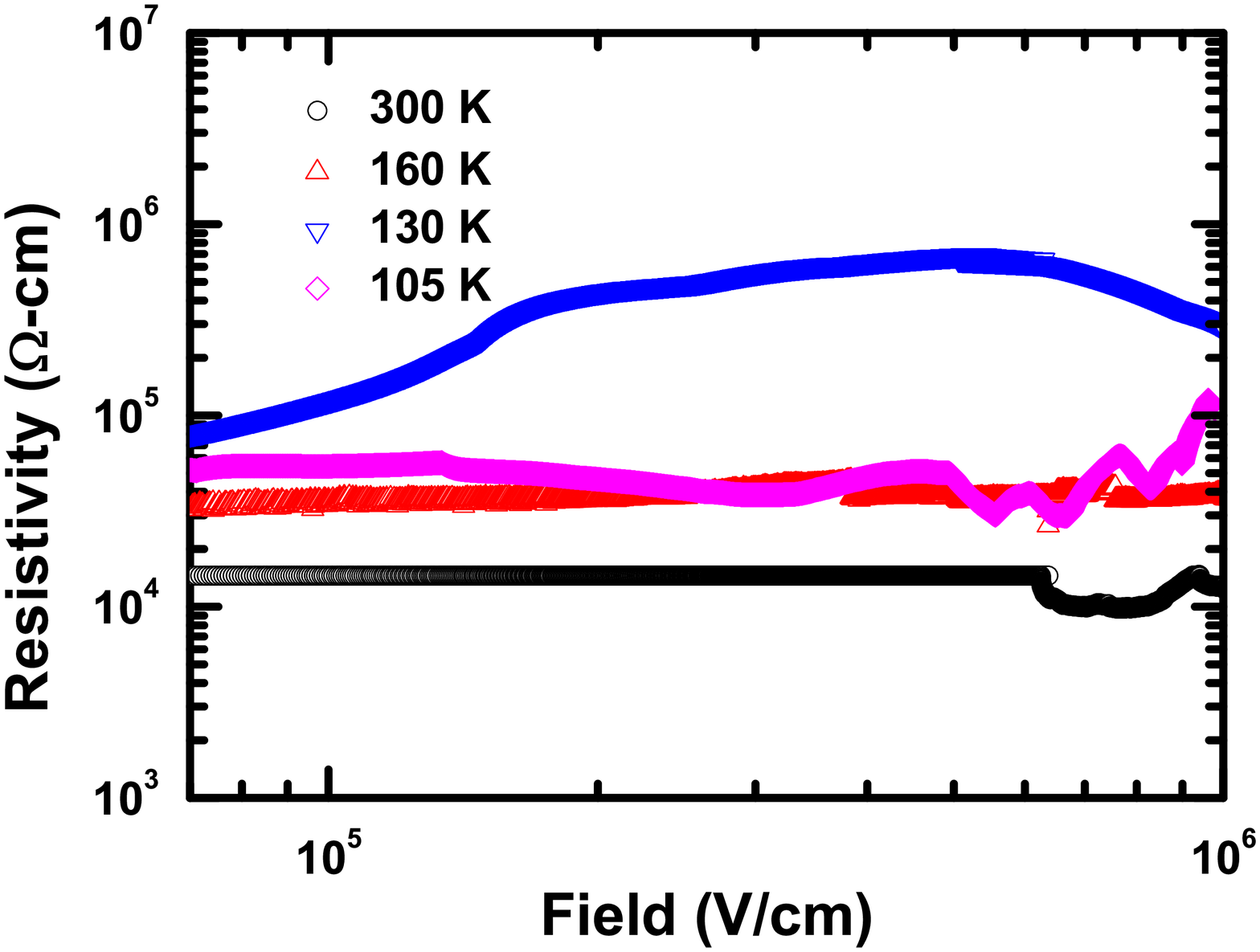}}
    \subfigure[]{\includegraphics[scale=0.20]{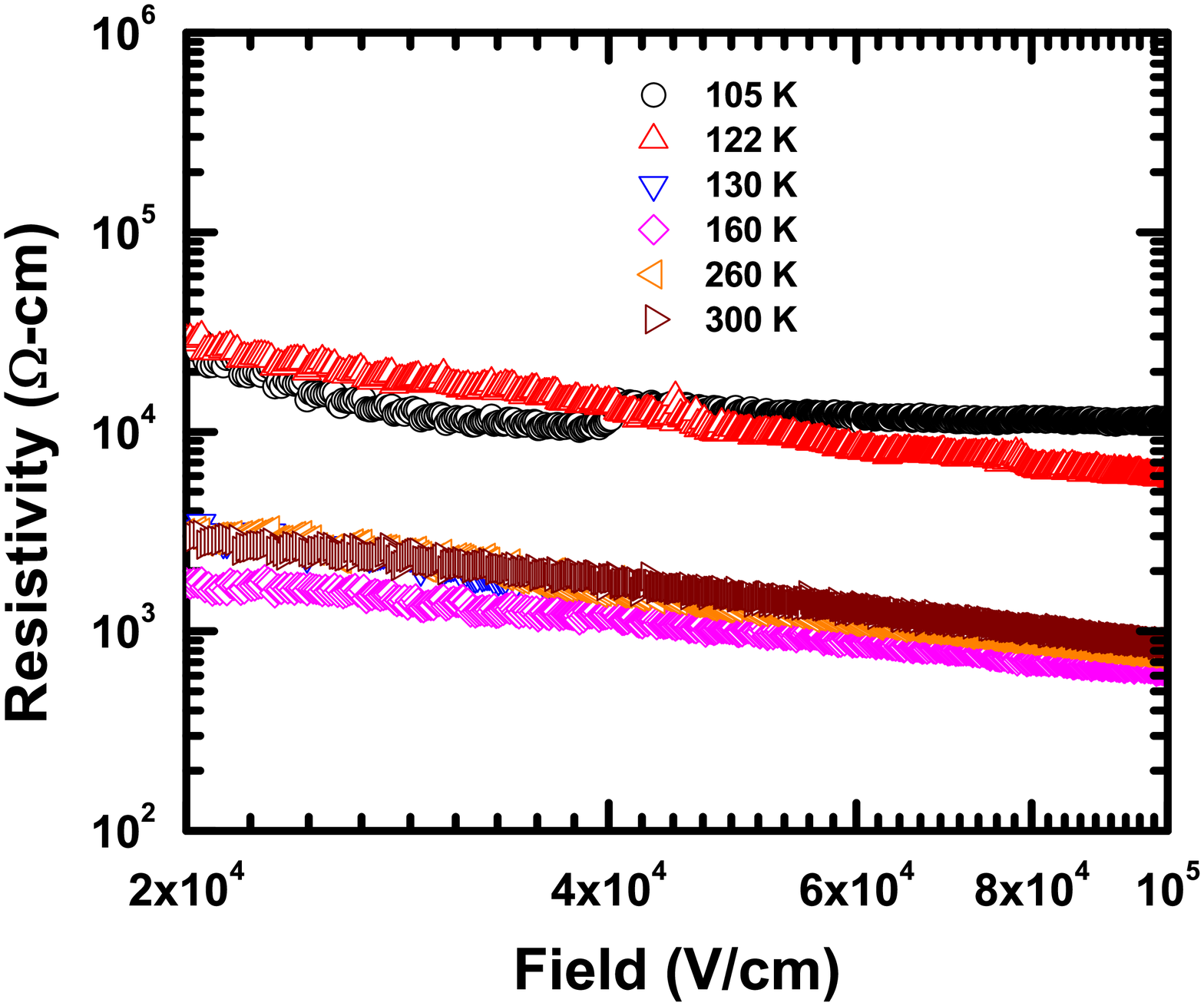}}
    \subfigure[]{\includegraphics[scale=0.22]{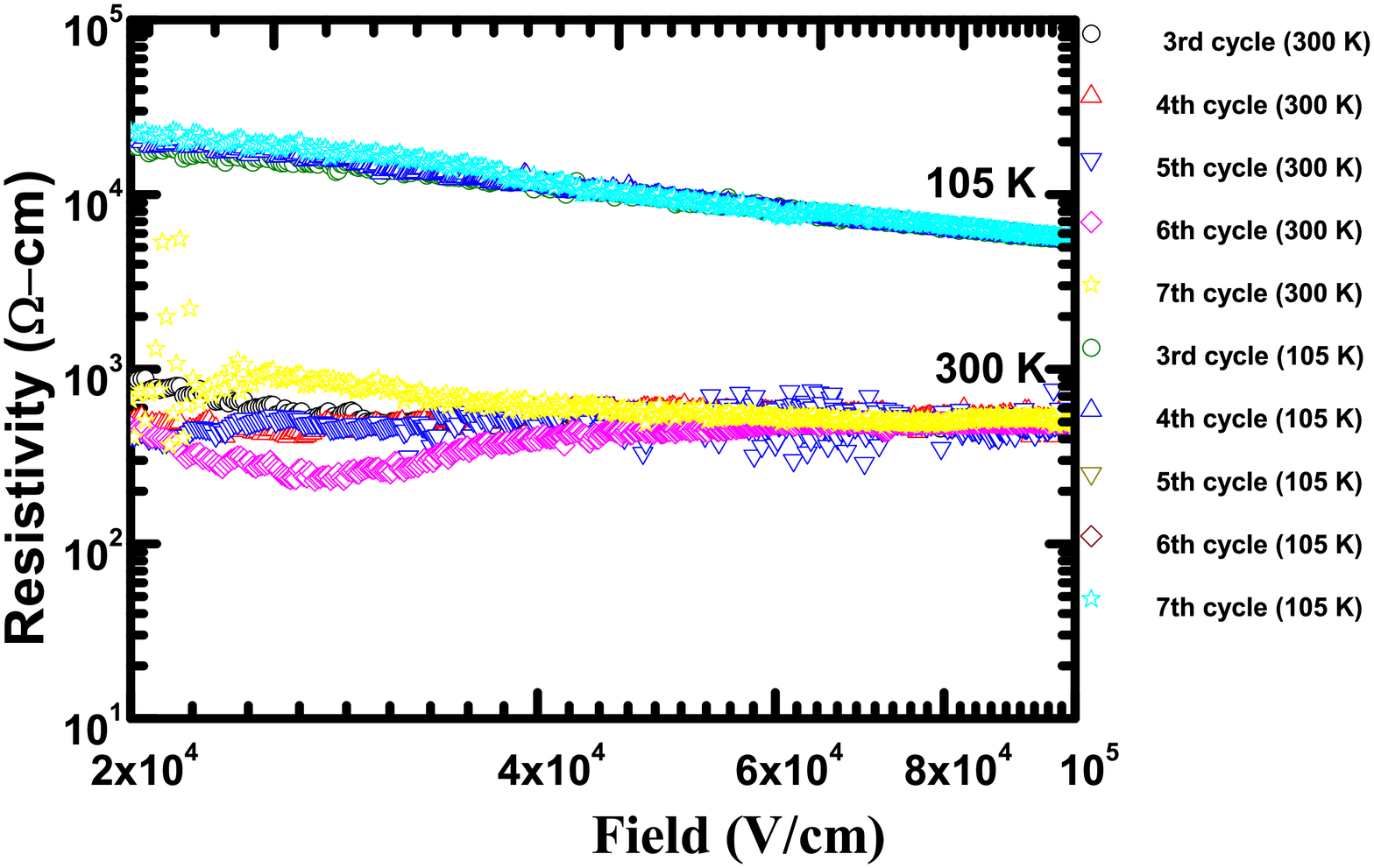}}
    \subfigure[]{\includegraphics[scale=0.20]{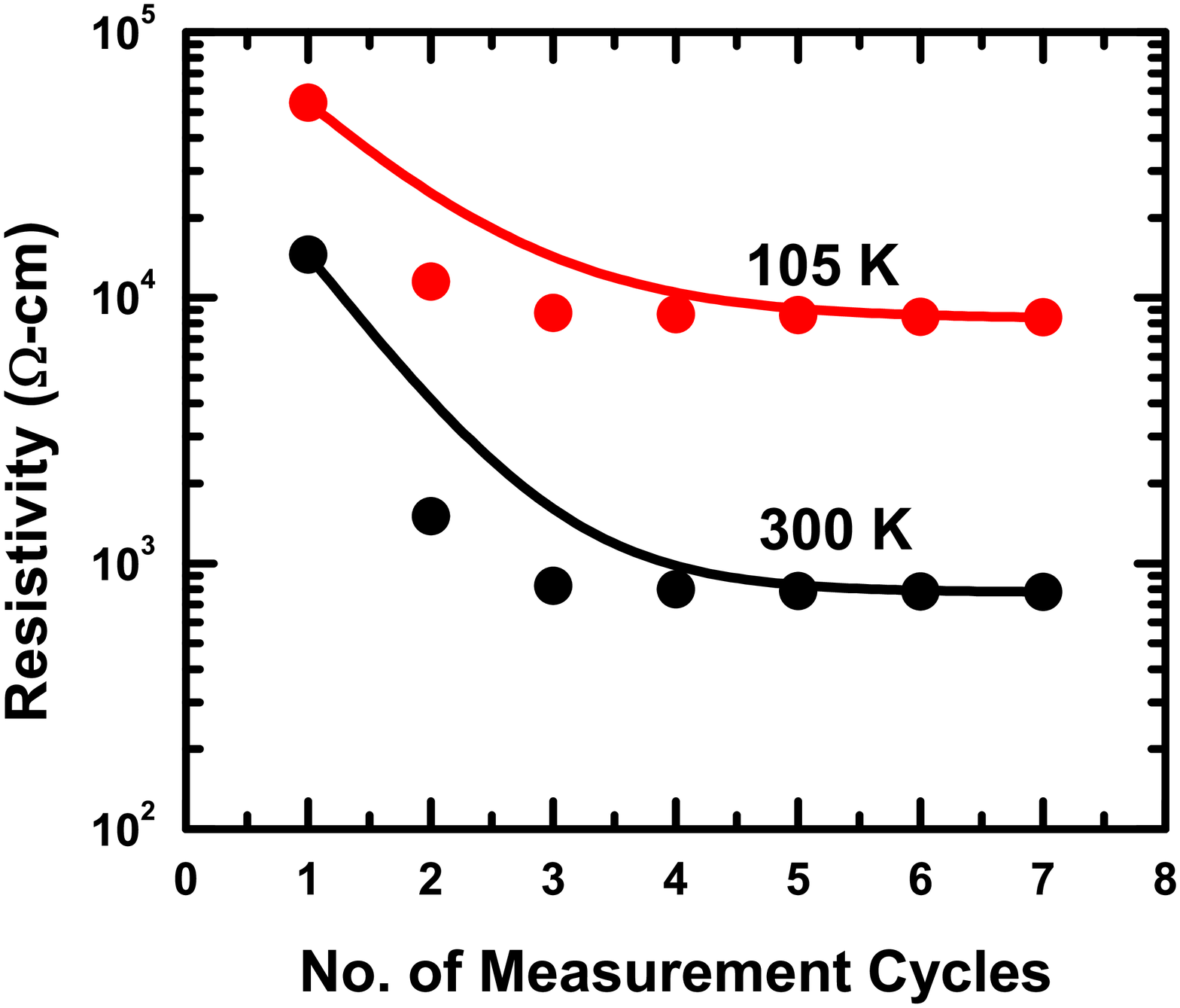}} 
  \end{center}
  \caption{(color online) The dc resistivity-field characteristics at different temperatures within 80-300 K for the pristine sample in (a) the first cycle, (b) the second cycle, (c) for the third to seventh cycles at representative temperatures (105 and 300 K) across the range, (d) dependence of resistivity on the number of measurement cycles at 105 and 300 K.}
\end{figure*}

\section{Results and Discussion}
In figure 2, we show the dc resistivity-field characteristics at different temperatures within the range 80-300 K. The sequence of measurement was as follows: the pristine sample was first taken down to 80 K from room temperature and during this first cycle, current-voltage characteristics were measured at several temperatures. The results obtained are shown in figure 2(a). The patterns are nearly flat within the range zero to $\sim$1.0 MV/cm. Only the data corresponding to $\sim$130 K show a slight rise in resistivity with the increase in field. Although this result apparently signifies a stable behavior, a second cycling through the same temperature and field range reveals a drastically different behavior. In figure 2(b), we show the resistivity-field characteristics observed in the second cycle. The resistivity values observed across the applied field range differ sharply from what has been measured in the first cycle. The third cycling offers a behavior similar to the second cycle albeit with a slightly different resistivity value. The difference between the data observed in second and third cycles is small. The difference progressively reduces as the measurement is repeated and finally becomes negligible beyond a certain number of cycles. This is, in fact, the central result of this paper. In figure 2(c), we show the patterns observed in third to seventh cycles at two representative temperatures within the range: $\sim$105 and $\sim$300 K. In order to illustrate the training effect more clearly, in figure 2(d) we plot the variation of the resistivity at 105 and 300 K as a function of number of measurement cycles. We found that the variation of resistivity from cycle to cycle follows nearly an exponential pattern (solid lines in figure 2(d)). The resistivity stabilizes via an activated defect hopping process (electromigration) over an energy barrier. The kinetics of such a process normally follow an exponential pattern with the driving force such as electric field and/or temperature. Since the cycle dependence of resistivity is influenced by the kinetics of the defect hopping process and reflects a crossover from unstable to a stable state via such kinetics, it also turns out to be following a nearly exponential pattern. In all these measurements in different cycles we have restricted ourselves to a low bias field regime where the current-voltage characteristics are either strictly linear or nearly linear and no switching occurs. As a result, while in the first cycle we could cover a field range of $\sim$10$^6$ V/cm, in the subsequent cycles we had to restrict the measurements within $\sim$10$^5$ V/cm only. We have measured the current-voltage characteristics with positive bias only at each temperature. Since we were mostly interested in exploring the training effect noticed during dc resistivity versus temperature measurement under a constant current and could anticipate that irreversible microstructural changes might lead to such a phenomenon, we did not explore the full linear and nonlinear regime of the current-voltage characteristics as well as the entire hysteresis loop with positive and negative bias. The influence of ac bias with changing polarity on electromigration driven damage has earlier been studied by others in which it has been shown that the change in polarity leads to a healing effect \cite{Tao}. The lifetime of the sample increases, as a result, under ac bias. We observe in our case that dc bias too is giving rise to a healing effect to the stress generated within the nanoscale sample during synthesis. It is important to point out here that the sample does itself become trained even in the first cycle of the measurement when the electric field is applied for the first time on a virgin sample. However, the training is not complete in the first cycle. As a result, change in the resistivity continues even in the subsequent cycles. The training is only considered to be complete when it stabilizes to such an extent that no significant change could be noticed in the data in between two subsequent cycles. The entire set of behavior is generic to all the nanowires/chains examined for this work although the pattern of variation of the resistivity with the number of measurement cycles may differ a bit. We have repeated this entire sequence of measurements on quite a few samples to see the training effect. In the supplementary document \cite{supplementary}, we provide the results obtained on three other samples patterned either by electron beam lithography or by FIB. We have given the data obtained in the first and final cycles in those cases. It is important to mention here that these features originate neither from the tungsten wire deposited by FIB as electrodes nor from the SiO$_2$/Si substrate on which the sample and electrodes have been deposited. In the supplementary document \cite{supplementary}, we give separate results of the measurements carried out on the substrate as well as on the electrodes.  

\begin{figure}[htp]
  \begin{center}
    \subfigure[]{\includegraphics[scale=0.35]{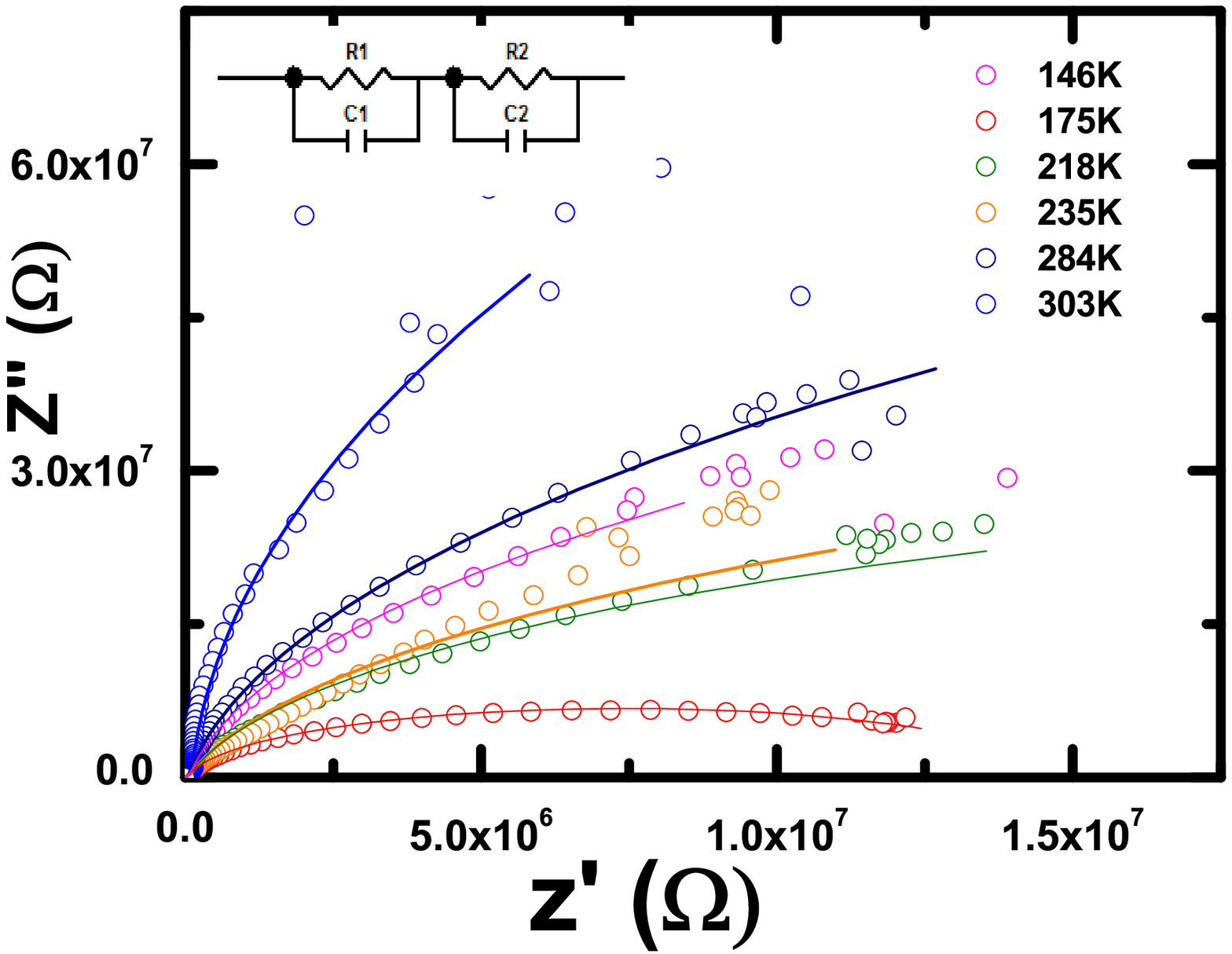}}
    \subfigure[]{\includegraphics[scale=0.35]{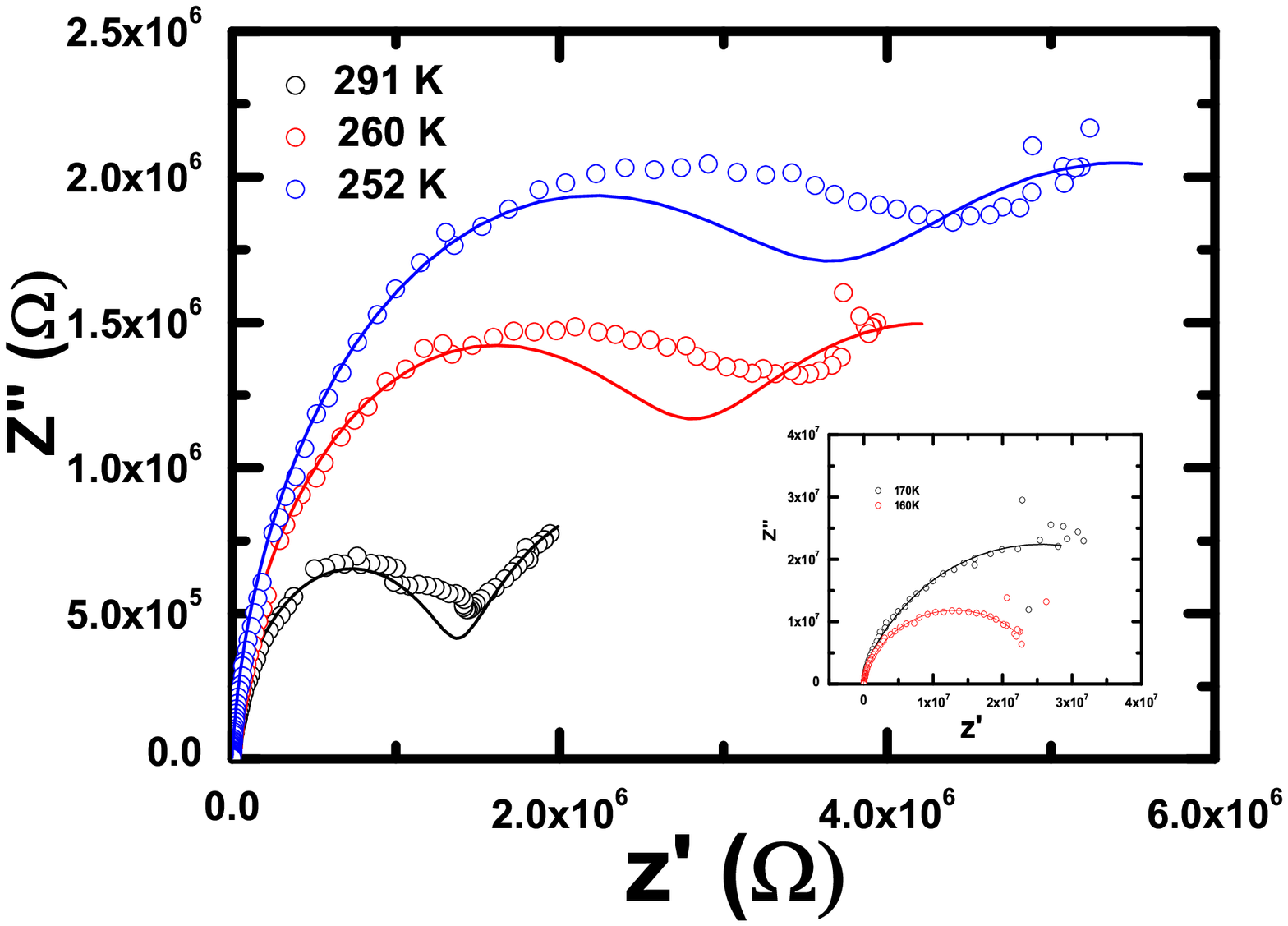}}
    \subfigure[]{\includegraphics[scale=0.25]{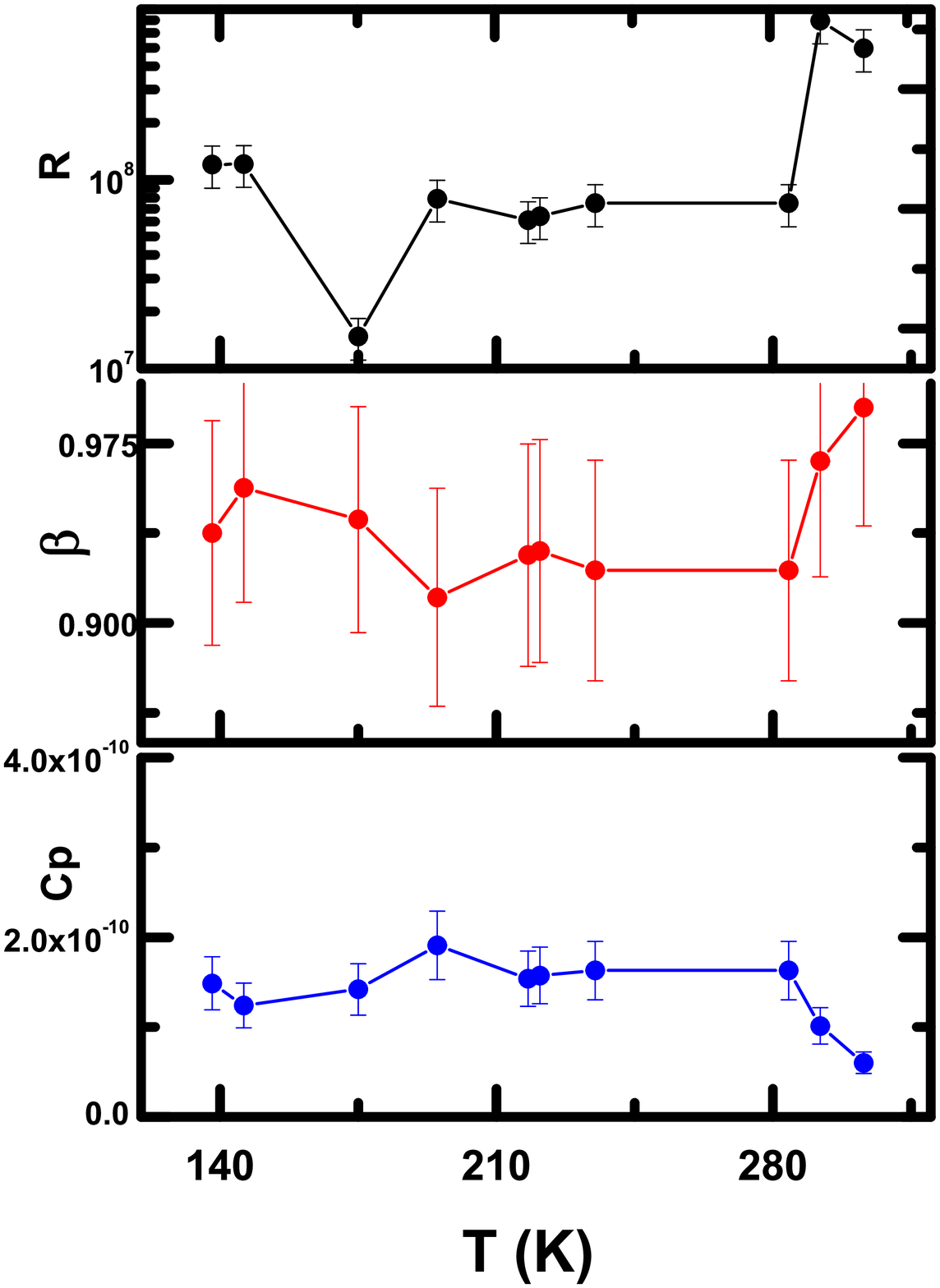}}
    \subfigure[]{\includegraphics[scale=0.20]{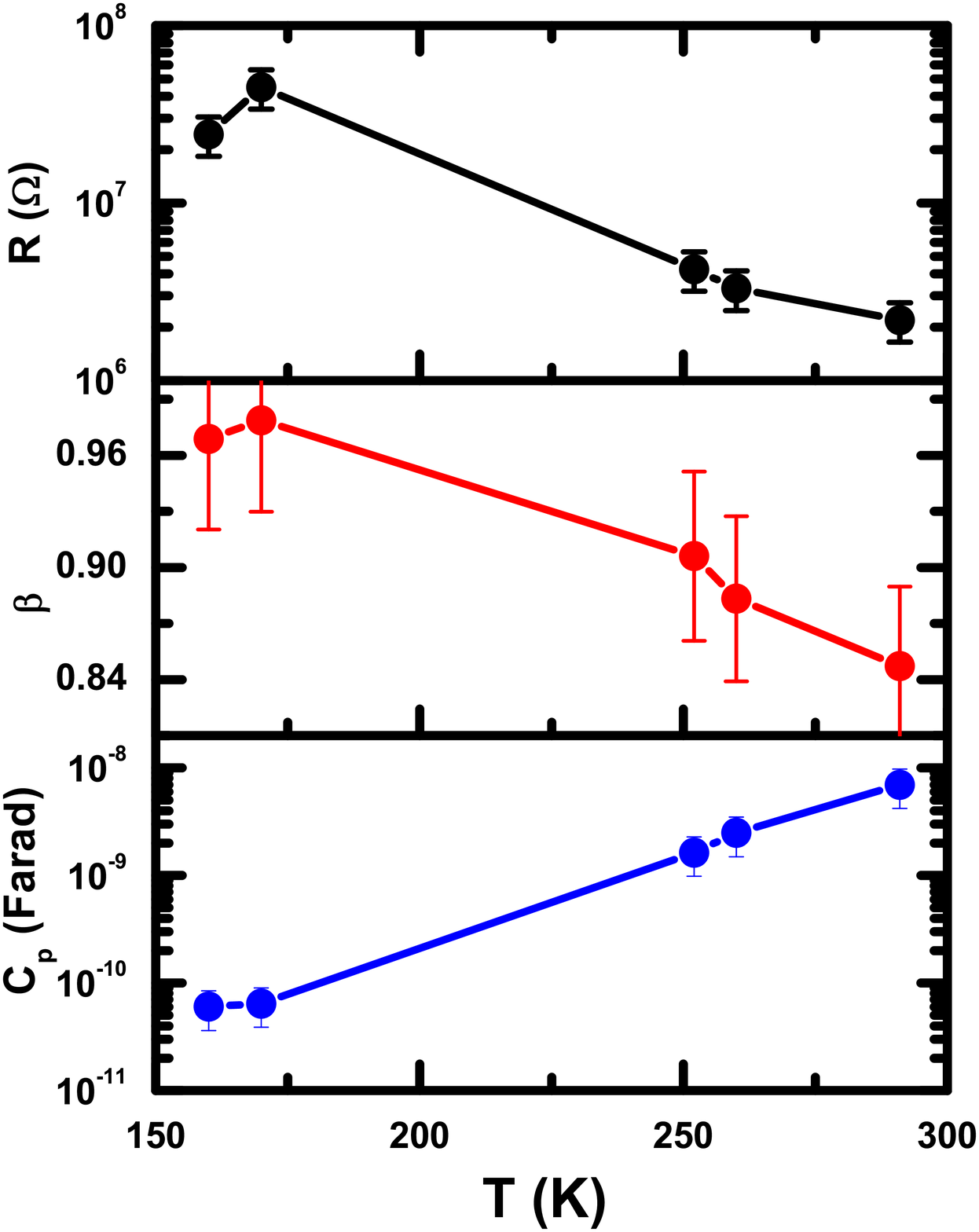}}
    \subfigure[]{\includegraphics[scale=0.25]{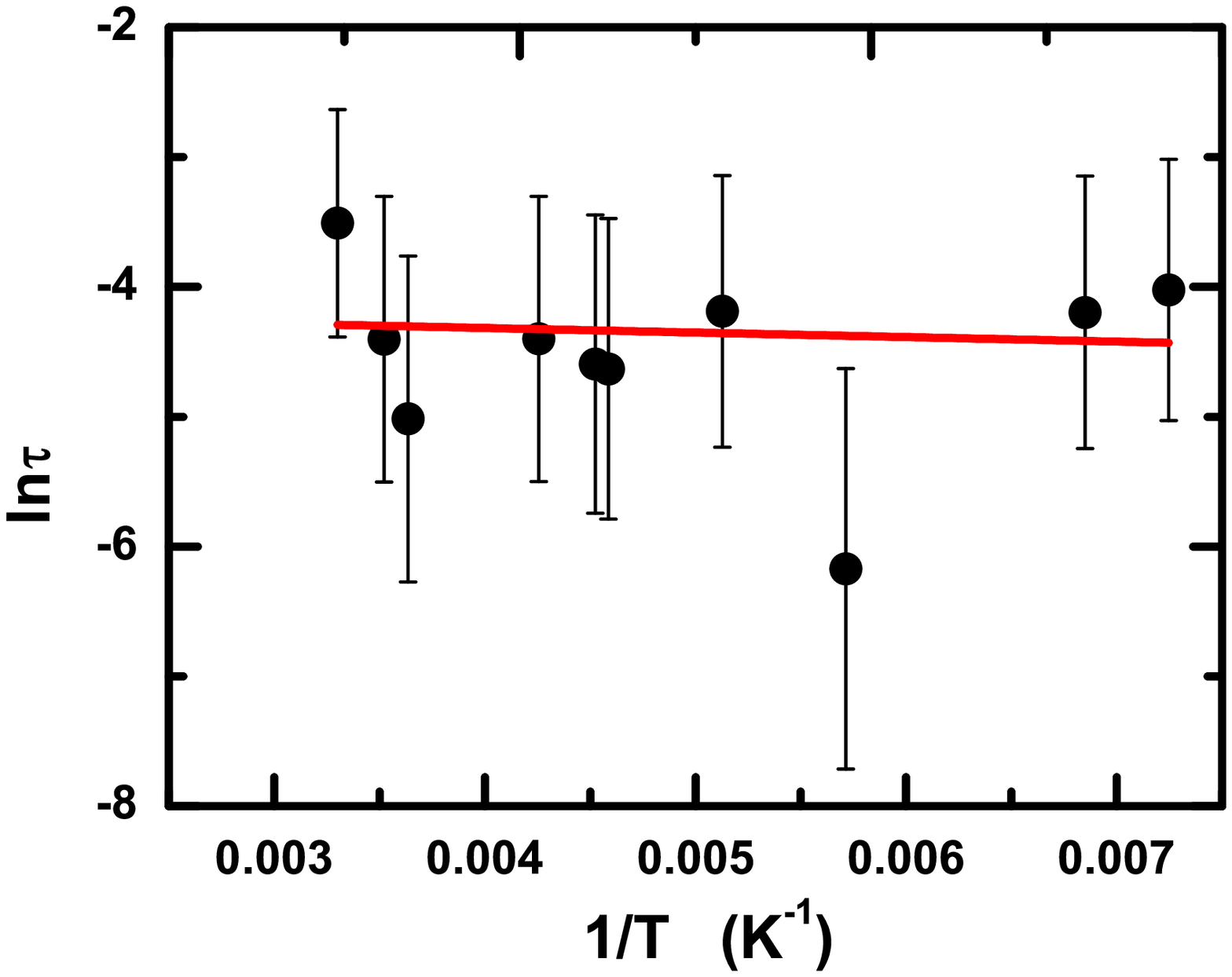}} 
    \subfigure[]{\includegraphics[scale=0.25]{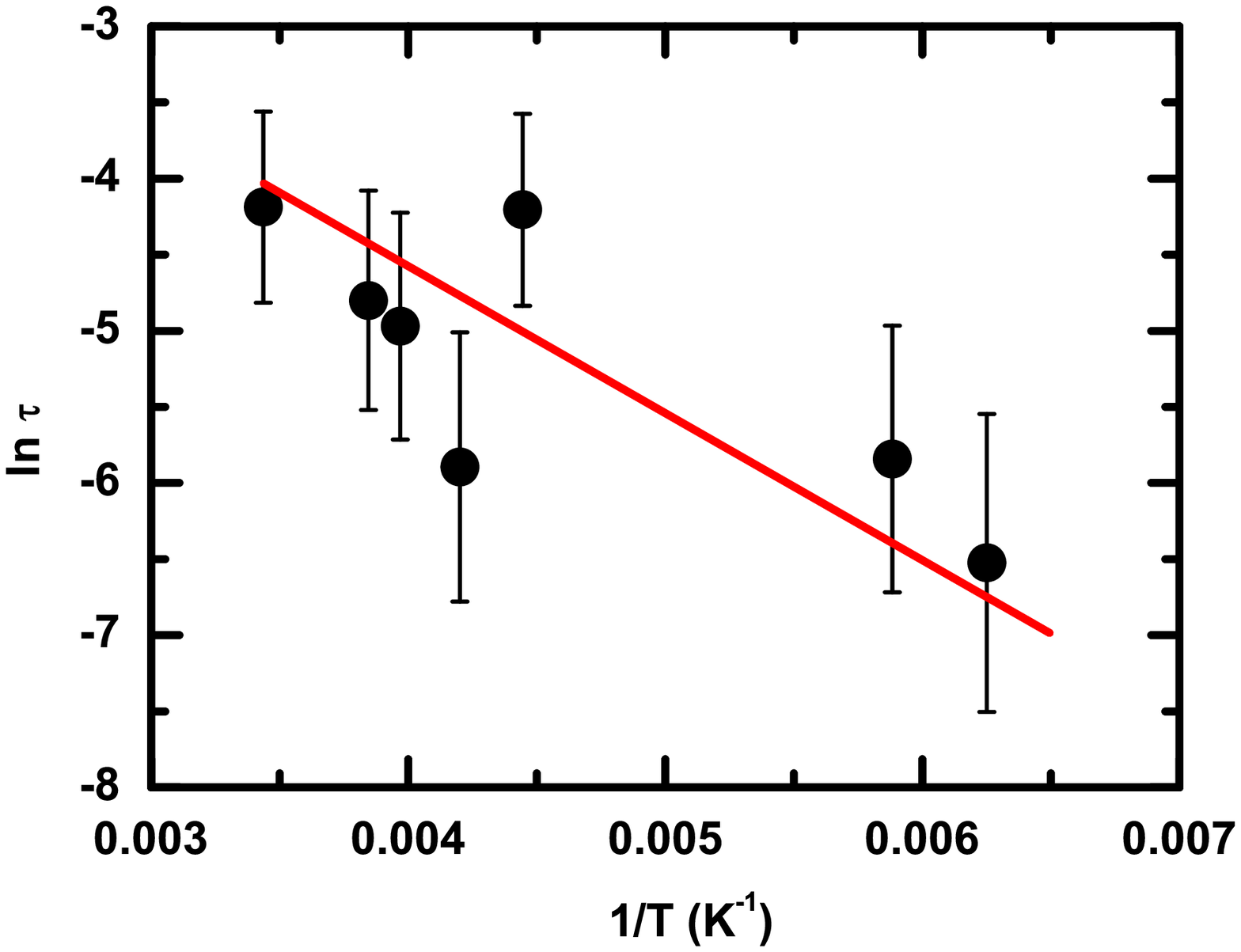}} 
  \end{center}
  \caption{(color online) The impedance spectroscopy data recorded on the virgin sample during the (a) first and (b) final cycle; inset of (a) shows the equivalent circuit diagram; the equivalent circuit parameters and $\beta$ are shown for the (c) first and (d) final cycle; the dielectric relaxation time constant ($\tau$) versus temperature plot is shown (e) in ln$\tau$-1/T frame for first cycle and (f) in ln$\tau$-1/T frame for final cycle; the change in the pattern of variation of $\tau$ with temperature from the first to seventh cycle is conspicuous.}
\end{figure}
 
Anticipating that the electromigration of defects might influence the dielectric relaxation characteristics, we have measured the dielectric relaxation patterns as well, both during the first and final cycle. In figure 3(a), we show the Nyquist plot of the data in Z"-Z' frame obtained from the measurements in first cycle. The relaxation of both the sample-electrode interface (Maxwell-Wagner) and bulk dielectric properties within the frequency range 100 Hz - 2 MHz does not take place in distinctly different frequency ranges. The fitting of the data, however, by an equivalent circuit model (figure 3(a), inset) of a series connection between two modules of resistors and capacitors connected in parallel, representing the bulk and electrode-sample interface regions, respectively, yields the sample resistance ($R$), capacitance ($C$), and the relaxation parameters ($\beta$). In figure 3(b), we show the Nyquist plot of the data obtained from the neasurement in final cycle. The inset of figure 3(b) shows the data corresponding to the lower temperatures. In this case, of course, the relaxation spectra of the sample-electrode interface and bulk dielectric properties occupy distinct regions within the entire frenquency range. The $R$, $C$, and $\beta$ obtained from the fitting of the equivalent circuit model with the experimentally obtained dielectric relaxation spectra from first and final cycles are plotted as a function of temperature in figures 3(c) and 3(d), respectively. The parameter $\beta$ signifies the broadening of the relaxation timescale. Quite interestingly while all the parameters such as $R$, $C$, and $\beta$ turn out to be nearly temperature independent during relaxation in the first cycle, one observes a clear temperature dependence - an increase in $C$ for a decrease in $R$ and $\beta$ with temperature - in the final cycle. This difference in the pattern signifies a distinct change in the dielectric relaxation characteristics between first and the final cycle. Figure 3(e) shows the relaxation time constant $\tau$ (=$R.C$) versus temperature in the ln$\tau$-1/T frame; $\tau$ is found to be decreasing very weakly with the decrease in temperature following nearly the $\tau$ $\sim$ e$^{-E_0/k_B.T}$ pattern, which indicates an unusual trend of a slowing down of relaxation kinetics with the rise in temperature. This is in complete contrast to the normal relaxation process in any dielectric system where a rise in temperature gives rise to faster relaxation and hence a shift in the relaxation time constant toward lower value. Such a trend has been observed in the case of a ferroelectric Ba(Ti,Zr)O$_3$ system within a small temperature region around the ferroelectric Curie point $T_C$ \cite{Itoh} or in the case of (KBr)$_{1-x}$(KCN)$_x$ (0.05$\le$x$\le$0.12) below 1 K for the boderline region of the compositions between crystalline system with isolated defects and an orientational glass \cite{Baier}. This anamalous feature in the present case indicates that indeed the relaxation here is taking place predominantly from the defect migration kinetics. As the temperature is raised, the clustered defects in the non-equilibrium configuration rearrange themselves into a stable equilibrium configuration leading to a slowing down of the migration kinetics and hence relaxation time constant. Interestingly, such a kinetics is observed not just in the first cycle but in the final cycle as well. In figure 3(f), the relaxation time constant corresponding to the final cycle has been shown in the ln$\tau$-1/T frame. The difference in the $\tau$-T patterns (figures 3(e) and (f)) between the data obtained in first and seventh cycles is readily apparent. The $\tau$-T pattern appears to be following nearly a $\tau$ $\sim$ e$^{-E_0/k_B.T}$ relation in the seventh cyle. While the activation energy E$_0$ turns out to be $\sim$3 meV in the first cycle, it increases to $\sim$56 meV in the final cycle. This shows that the defect migration kinetics continues to slow down with the cycling. This is akin to the kinetics of glass transition \cite{Santen}. 

\begin{figure}[!h]
  \begin{center}
    \includegraphics[scale=0.20]{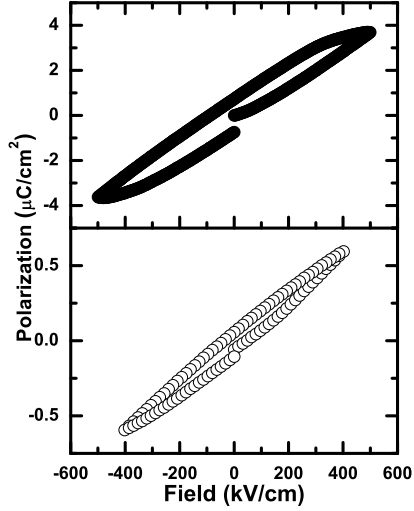} 
    \end{center}
   \caption{ The polarization (P) versus electric field (E) loop measured at room temperature on nanoscale BiFeO$_3$ in (bottom) the virgin state and in (top) the stabilized state. }
\end{figure}

\begin{figure}[!h]
 \begin{center}
    \includegraphics[scale=0.75]{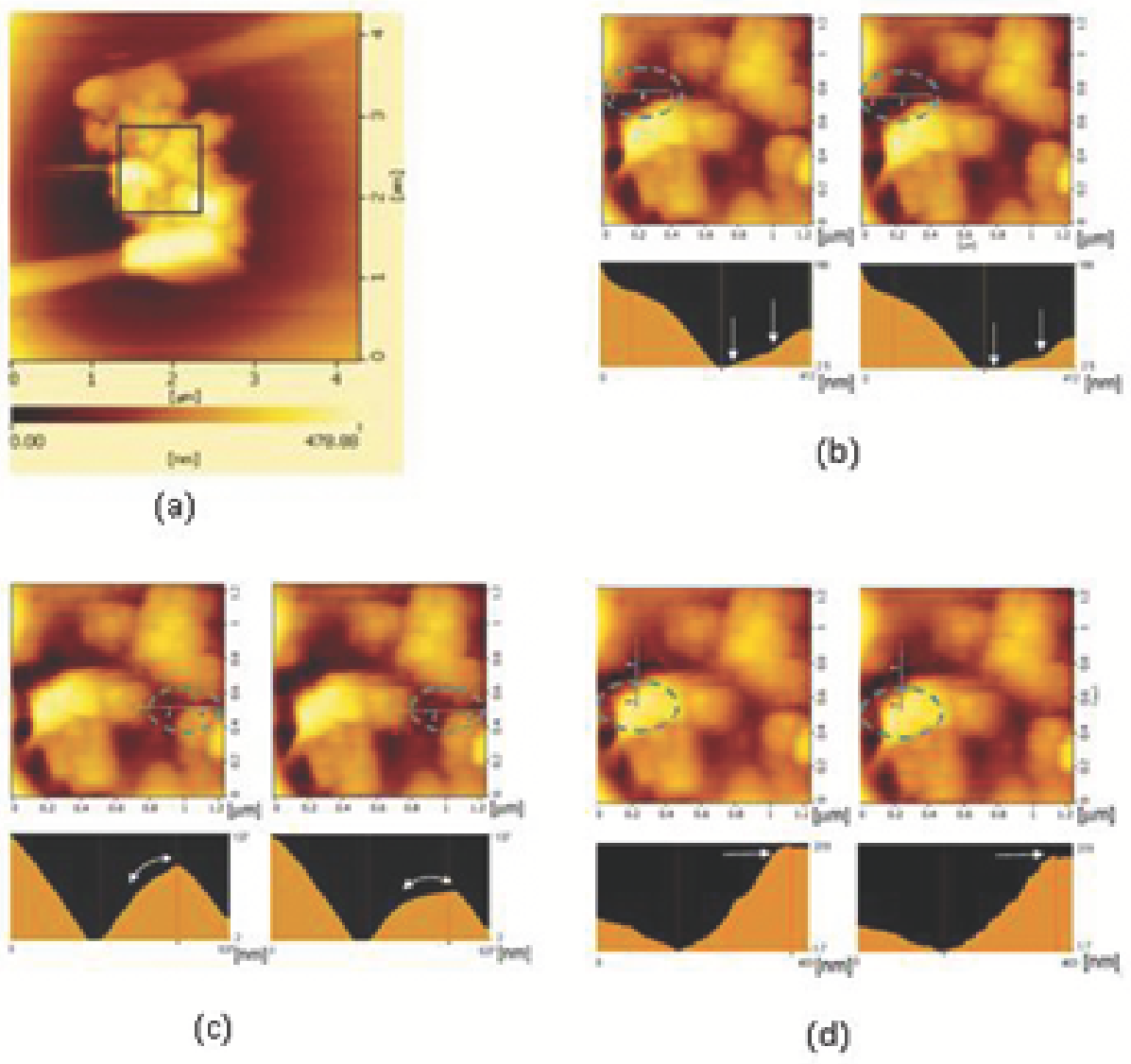} 
    \end{center}
  \caption{(a) The AFM image of a pristine nanochain of BiFeO$_3$ connected by FIB deposited tungsten wires before any measurement; the scale of the image is 4.5 $\times$ 4.5 $\mu$m$^2$; a portion of this full image of size 1.2 $\times$ 1.2 $\mu$m$^2$ is enlarged to illustrate the impact of electric field on the microstructure; (b)-(d) linescans have been taken at different positions on that portion of the full image shown in (a) to demonstrate the change in the microstructure quantitatively between zero and $\sim$0.125 MV/cm field at room temperature captured $\textit{in situ}$; three pairs of images are shown with each pair consisting of images under zero and $\sim$0.125 MV/cm field; the left image of any pair shows the microstructure under zero field while the right image shows that under $\sim$0.125 MV/cm field; alongside each image, the topographical profiles with quantitative measurements corresponding to the line scan of the image are shown at the bottom of each pair; while the changes observed in the images (b) and (d) are a bit small, that observed in image (c) is the largest and quite conspicuous. The portions of the topographical profiles which signify the changes in the microstructure have been marked by arrows.  }
\end{figure}

Since BiFeO$_3$ is an important multiferroic system, we also examined how the ferroelectric polarization of the nanoscale BiFeO$_3$ changes from a virgin to a completely trained sample. In figure 4, we show the polarization (P)-electric field (E) loop measured on a nanoscale BiFeO$_3$ in the virgin as well as in the stabilized state. While the loop measured in the virgin state does not reflect switching of polar domains, the one measured in the trained state does signify domain switching and closely resembles the ferrolectric polarization loop measured on nanoscale sample of other systems \cite{Tian}. This could be because randomly distributed large scale defect structure possibly pins the domains and prevents switching in the virgin state. In the stabilized state, even distribution of the defects appears to be allowing the switching of polar domains under sweeping electric field. This result shows that the training is necessary before any direct electrical measurement is carried out on nanoscale BiFeO$_3$ in order to ascertain its characteristic properties.     

Direct evidence of large scale defect migration has been gathered from $\textit{in situ}$ AFM images under a bias field at room temperature. The FIB-patterned nanoscale BiFeO$_3$ sample has been biased at room temperature with a dc voltage across a range 0-10 V which corresponds to a field range of 0-0.125 MV/cm. Initially, the AFM images in the pristine state (under zero field) have been captured in contact mode. The electric field was then applied by a sourcemeter via the FIB deposited connectors through the sample and $\textit{in situ}$ AFM images have been captured again in contact mode. In figure 5, we show the AFM images of a pristine nanochain under both zero and $\sim$0.125 MV/cm field at room temperature. Figure 5(a) shows the entire device with nanoscale sample and FIB deposited electrodes in the pristine state. A portion of that microstructure (boxed part of the image in figure 5(a)) is blown up in the subsequent images to demonstrate the change in microstructure under electric field at three different regions. There are three pairs of images; figure 5(b), 5(c), and 5(d). In each pair, the left image shows microstructure under zero field and the right one shows that under a bias field of $\sim$0.125 MV/cm. The grain-grain boundary microstructure - especially, the curvatures of the grain surfaces and interfaces - change under the field. The topographical profiles, scanned at different region of the microstructure along the lines marked in each pair of images and shown at the bottom of the corresponding image, depict conspicuous modifications resulting from migration of defects under the bias field. These topographical profiles show that the height of the top surface of the microstructure changes from 134 nm to 127 nm in the top two images; from 310 nm to 285 nm in the middle two images and from 94 nm to 61 nm in the bottom two images. Apart from changes in the height, it is possible to note subtler changes as well such as depletion in one region and swelling in another. Those regions are marked by circles in the topographical profiles. These could signify large scale defect migration from one region to another. These changes are irreversible and, therefore, illustrate the training effect on the microstructure. This, in turn, trains the dc electrical properties shown in figure 2. It is important to mention here that this large scale defect migration and, thereby, noticeable change in the topographical profiles across the grain-grain boundary structures, occurs when electric field is applied on a virgin sample, i.e., in the first training cycle. Obviously, in this cycle the change in the electrical resistance too is maximum. In the subsequent cycles, the change in the electrical resistance is small and the defect migration and consequent change in the microstructure too is rather subtle and not quite detectable by the AFM. The defect structure, of course, eventually stabilizes with drastic slow down in the migration kinetics as noticed from the dielectric relaxation characteristics. Li $\textit{et al}$. \cite{Li1}, on the contrary, report severe impact of electromigration leading to splitting of the GaN nanowires under prolonged current flow.    

As pointed out earlier, this behavior is generic to all the nanoscale structures examined for this work. In the pristine state, the defects in non-equilibrium state lie in shallow barriers and undergo a large scale migration under combined influence of Joule heating and electric field leading to a phase transtion from the initial metastable phase to a stable one \cite{Tomar}. In the stabilized state, the defects form a stable network with a drastically slowed down migration kinetics. The difference in activation energy of dielectric relaxation reflects this switch in the migration kinetics. It appears that all such structures require a carefully chosen thermal and electric field cycling treatment in order to reach the stable state.  

\section{Summary}
In summary, we show in a nanochain of BiFeO$_3$, from dc current-voltage and  dielectric relaxation characteristics as well as from $\textit{in situ}$ AFM images, a training effect on the electrical properties as a result of a large scale migration of defects. We conclude that such a training effect is ubiquitous in nanoscale structures of oxides and it needs to be addressed when preparing them for applications in nanoelectronics or nanospintronics.

$\textbf{ACKNOWLEDGMENTS}$

This work has been supported by CSIR-NSFC joint program on nanoscale multiferroics, the National Natural Science Foundation of China (Grant Nos 50825206 and 91123004), and Outstanding Technical Talent Program of the Chinese Academy of Sciences. One of authors (S.G.) acknowledges support from a Research Associateship of CSIR.

\end{document}